\documentclass{article}
\usepackage{amsmath}
\usepackage{xcolor}
\usepackage{graphicx} 
\usepackage{dsfont}
\newcommand{\ket}[1]{\left | #1 \right\rangle}

\newcommand{\identity}{\mathds{1}}
\begin{document}
\title{Tests of constructor theory}
\author{Chiara Marletto, David Deutsch, Vlatko Vedral}
\maketitle
\begin{abstract}
Constructor theory is a proposal to extend quantum information theory beyond both quantum theory and computation, to cover more general machines than programmable computers -- called constructors. It consists of newly conjectured physical principles that can be expressed as constraints on what tasks are possible, what are impossible, and why. These principles also determine the repertoire of the universal constructor, which is a programmable machine {that can perform all physically possible tasks}. The principles of constructor theory have novel physical content that supplements current dynamical laws, leading to new predictions for experimental tests. In this paper, we review the main experimental proposals to test the principles of constructor theory and discuss their implications for existing theories of physics and for their successors.
\end{abstract}

\section{Introduction}

The advent of the quantum theory of computation has brought significant technological developments, as well as (and perhaps more importantly) profound theoretical advances. It may come as a surprise, but one of the primary aims of the theory of quantum computation, when it was first proposed in the 1980s, was to deepen our understanding of quantum physics. For instance, entanglement measures, which illuminated the solution to the measurement problem and clarified how observations work within unitary quantum theory, are pillars of the modern understanding of quantum theory. Remarkably, those conceptual and theoretical advances in fundamental quantum physics are direct offspring of merging quantum theory with information theory. The technological progress ensuing from the search for universal quantum computers is, considered from this angle, only a side effect. However, it was only when the theory of quantum computation resulted in actual experimentally relevant protocols that it became accepted as a fully fledged field within physics. Indeed, in quantum information, abstract and fundamental theoretical ideas (for instance, the theory of the universal quantum computer) very soon became relevant for technological implementations, experimental demonstrations, and experimental tests. This story is a good example of how theoretical physics, experimental physics and technological development fruitfully complement each other. 
In this review, we explore the experimental implications of constructor theory -- a programme that has been proposed to extend the quantum theory of computation in a number of promising directions. 
Constructor theory's fundamental motivations, vision and scope were proposed by D. Deutsch in \cite{DEU}.  C. Marletto and D. Deutsch applied it first to information theory \cite{DEUMA, THE}; C. Marletto then drove a programme to deliver applications of this approach in thermodynamics \cite{MAT1, MAT2, VIOMA}, the physics of life \cite{THE, MAL}, and the physics of hybrid systems \cite{MAP, MAVE17c, MAVE17d, MAVE20}. This has led to interesting proposals for experiments and to actual experimental demonstrations, \cite{MAVE17d, MAVE17a, MAVE17b, MAGE}. {In this review we will focus on the experiments of constructor theory, and summarise its basic theoretical ideas in Section 2 and in the Appendix.}

One way in which constructor theory extends quantum information theory is that it can provide the theory of more general machines than computers, called `programmable constructors' (the theory takes its name from these machines). First proposed and studied by J. von Neumann, \cite{von}, constructors generalise computers, because they can perform all sorts of physical tasks, not just computations: {for instance, enzymes, catalysis, and heat engines are instances of (approximate) constructors}. Unlike von Neumann's theory of the universal constructor, which never went beyond the domain of cellular automata, constructor theory is a proposal for a novel theory of physics, so it is intended to supplement current physical laws. In addition, constructor theory emancipates information theory from quantum theory itself, as it does not rely on its dynamical laws or its formalism. This feature is essential to make the quantum theory of information `portable', so that its results may be applied also to the successor of quantum theory, which may not be expressed in the same formalism as its predecessor. 

Another way in which constructor theory goes beyond the quantum theory of computation is that it provides a new theory of physics. It supplements existing dynamical laws and allows us to make predictions in areas where dynamical laws are unable to make progress at present. For instance, it can express laws about entities which are traditionally regarded as `macroscopic' (such as information carrying media, computers, and heat engines), without using approximation schemes or scale-dependent statements. The main reason for this is that constructor theory's principles are expressed in terms of what transformations (called `tasks') are possible and what are impossible, and why. The concept of possibility and impossibility are primitive in constructor theory, but they can be modeled within dynamical laws (see Section 2 and the Appendix for more details). In short, this can be done by defining a constructor for a task as a physical system that can perform the task working in a cycle (i.e., retaining the ability to perform it again). Then, a task is possible if it can be performed by a constructor to arbitrarily high accuracy; it is impossible otherwise. By abstracting away the constructor from the laws, and only referring to possible/impossible tasks, constructor theory can express laws in a scale-independent way even for entities traditionally regarded as emergent, like information-bearing systems or living entities. Hence constructor theory has the potential to integrate quantum information theory, thermodynamics, and the physics of life in a single corpus of laws that belong to fundamental physics. 
Assuming that a universal constructor is allowed in our universe, constructor theory will also deliver the theory of the universal constructor \cite{DEU25}. Indeed, the study of the universal constructor's repertoire (i.e., the set of tasks it can perform) is isomorphic to the whole of physics. Studying the universal constructor and its repertoire will deepen our understanding of the foundations of physics, much as the theory of the universal quantum computer deepened our understanding of quantum theory.  

The principles of constructor theory are meta-laws: they are obeyed by some dynamical laws (called `subsidiary theories'), and violated by others. Subsidiary theories are not part of constructor theory: the full explanation of a given physical situation is given by the principles of constructor theory and by the compatible subsidiary theories. The principles are formulated in a scale- and dynamics-independent way, so they constrain a broad class of subsidiary theories; they do not refer to constructors but rather to the possibility or impossibility of certain tasks.
{If constructor theory is true, its principles supplement the dynamical laws and the initial condition theories, which become emergent consequences of the principles; just as, for example, time and dynamics are emergent in the Page-Wootters construction in a timeless universe, \cite{PAGE}}. The 
compatibility between existing dynamical laws and the theory of clocks and time in constructor theory is 
discussed in \cite{DEUMA25}.

\subsection{Testing constructor theory}
Testing the principles of constructor theory works in the same way as testing other principles of physics, such as the conservation of energy or the second law. One must test them not by direct falsification of their predictions, but by testing the subsidiary theories that comply with them, or by showing that given tasks that are ruled out by them are, in fact, possible, (as explained in detail in \cite{DEUTEST} and the references herein).

Recent developments have shown that it is already possible to test some of the principles of constructor theory, \cite{MAVE17d, MAVE17a, MAVE17b, MAGE}. In this paper, we review the experimental tests of constructor theory that have been proposed and performed so far, and discuss possible future ones, along the following lines:

\begin{itemize}
\item{\bf Constructor Theory of Information}\\
The constructor theory of information unifies classical and quantum theories of information, as well as other post-quantum theories, \cite{DEUMA, MAP}. It also proposes physical principles to express the regularities that allow for information to be embodied in physical systems. This constitutes the ideal tool to describe concepts of observables without using quantum theory's dynamics or formalism, hence allowing us to design new tests of hybrid quantum-classical systems. C. Marletto and V. Vedral have applied this theory to the physics of hybrid systems, where a quantum probe interacts with a system whose dynamics is partially unknown and may deviate from quantum theory itself, \cite{MAVE17d, MAVE17a, MAVE17b}. This has resulted in a novel class of witnesses of non-classicality, which can be applied to design tests of quantum features in a variety of contexts, from quantum gravity to quantum biology. We shall review these tests in detail, highlighting how they are also an indirect test of the constructor theory's principles of information, notably, the interoperability of information and locality. 
\item{\bf Constructor Theory of Thermodynamics}\\
Constructor Theory offers a new foundation for thermodynamics: it can express the second law exactly, as a scale-independent, dynamics-independent principle, generalising existing axiomatic approaches, \cite{LIE}; it has also established an unexpected link between the conservation of energy and information theory. In particular, C. Marletto has defined thermodynamic work in scale-independent terms, by providing constraints on systems that can do work, expressed only in terms of possible/impossible tasks, and without using particular dynamical laws formalisms, coarse-graining or statistical mechanics methods, \cite{MAT1, MAT2}.
Experiments testing the fundamental limits of quantum heat engines, Maxwell's demon, and non-equilibrium thermodynamics could provide empirical evidence for or against the principles of constructor-theoretic thermodynamics. We review the experiments performed so far, which have demonstrated a novel kind of irreversibility arising from the laws of the constructor theory. This irreversibility, based on requiring that a task is possible, but its inverse task is not, was proven by M. Violaris and C. Marletto to be exactly compatible with time-reversal symmetric laws, \cite{VIOMA, MAGE}. 
\item{\bf Testing other aspects of constructor theory} \\
{This more speculative section of this review describes possible future tests of constructor theory, concentrating on the following areas.} The constructor theory of life, \cite{THE, MAL} investigates what physical principles are sufficient in order for highly accurate self-reproduction to be possible under our laws of physics, in a universe where the physical laws are not especially tailored for life. This work also unveils the necessary features of accurate self-reproducers, such as bacteria, demonstrating that they must contain a replicator (a DNA-like set of instructions, which can be copied) and a vehicle that serves as a programmable constructor to execute the self-reproduction. In passing, it also provides a physical definition of what constitutes ``knowledge" in a physical system, including in living organisms, as a particular kind of information that can act as a constructor to keep itself in existence. This work may lead to tests of how information in biological systems adheres to constructor-theoretic principles (e.g., Darwinian evolution as an irreversible information-processing transformation) and also of how the principles of constructor theory could be tested to study the transition from no life to early life. 
We shall also consider other possible avenues towards tests of constructor theory, such as how the derivation of a stochastic law which generalises the Born Rule in the constructor theory of information can lead to new tests of the Born Rule, \cite{MAP}. 
Finally, we shall speculate about possible tests related to the theory of the universal constructor, \cite{DEU25} and to the role of the principles of constructor theory as selectors of subsidiary theories.
\end{itemize}

\subsection{Fundamentals of constructor theory} 

In this section, we provide an informal summary of the motivations and the basic notions of Constructor Theory (for more formal details, refer to the appendix and to \cite{DEUMA, THE, MAP, MAT2}). 

\subsection{Motivations and rationale}

{We will now summarise the broad motivations behind constructor theory. Standard physical theories are formulated primarily in dynamical terms: they specify laws of motion and initial conditions, from which actual state evolutions are derived. This paradigm, albeit successful, does not naturally accommodate `counterfactual statements': statements about which physical transformations are possible or impossible independently of whether they occur. Such statements are key to concepts such as information, computation, controllability, measurement, and thermodynamic irreversibility. Consequently, in standard physical theories, these concepts appear only indirectly or as emergent features of underlying dynamics, rather than as fundamental physical notions. Moreover, many foundational principles in physics are inherently counterfactual — for example, the impossibility of perfect cloning, perpetual motion, or superluminal signalling — yet existing formalisms lack a unified language in which such impossibility statements are primary. Notions such as computation, information, and reliable construction cannot be fully reduced to differential equations or Hamiltonian evolution because they depend essentially on counterfactual properties — that alternative transformations would remain possible under varying circumstances. Constructor theory addresses these limitations by providing a common foundation for e.g. information theory, thermodynamics, and aspects of computational biology. We also note that extending current dynamical frameworks incrementally would not suffice to solve these issues, because the underlying explanatory mode remains unchanged. Finally, constructor theory can be considered as an extension of the quantum information theory, as discussed earlier in the introduction. While the current framework of quantum information theory is both operationally successful and internally coherent, it is rooted within quantum theory's formalism and therefore is not sufficiently robust. Constructor theory aims at expressing the formalism-independent aspects of quantum information theory. The reason for this is that several current approaches to quantum gravity and post-quantum frameworks suggest that core ingredients of standard quantum theory — including fixed Hilbert-space structure, subsystem factorisation, and even standard notions of time evolution — may only be approximate or emergent. In such settings, concepts such as entropy, entanglement, distinguishability, or channel capacity may require new formulations that do not presuppose the full machinery of ordinary quantum theory. Constructor theory's goal is to investigate which informational principles are genuinely structural and which depend on contingent features of the current formalism.
One possible outcome is that existing quantum-information-theoretic results can indeed be translated into a successor theory, much as classical information theory embeds naturally into quantum theory. However, carrying out such a translation presupposes identifying which aspects of the theory are dynamics-independent in the first place, which is what constructor theory aims to.}

\subsection{Basic notions}

{The core concept in constructor theory is the notion of a ``task." A task specifies a transformation in the form of an ordered pair of input and output ``attributes", representing physical features of systems on which a task can be performed, which are called ``substrates".  An {\sl attribute} ${\bf x}$ is a set of states of a substrate that share a common property $x$. States are defined by each subsidiary theory on a space which must be endowed with at least a topology.  In quantum theory, for example, the set of all quantum states of a qubit in which a given projector $\Pi$ is sharp with value $1$ constitutes an attribute. 
If ${\bf a}$ is an attribute of substrate ${\bf S_1}$ and ${\bf b}$ is an attribute of substrate ${\bf S_2}$, then the attribute ${\bf (a,b)}$ of the composite substrate ${\bf S_1}\oplus {\bf S_2}$ is defined as the set of all states in which ${\bf S_1}$ has attribute ${\bf a}$ and ${\bf S_2}$ has attribute ${\bf b}$. {In quantum theory, one can represent attributes with projectors.} Assuming a two-qubit Hilbert space} ${\cal H}_{ab}$, \begin{equation}
{\bf (a,b)}\doteq\{ \rho_{ab}\in {\cal H}_{ab}\;:\;{\rm Tr}\{\rho_{ab}\Pi_a\otimes \Pi_b \}=1\}\;
\end{equation} where $\Pi_a$ and $\Pi_b$ are projectors acting on each qubit's Hilbert space. 
According to Einstein's principle of locality, if a transformation acts solely on substrate ${\bf S_1}$, then only attribute ${\bf a}$ may change, while ${\bf b}$ remains unaffected, \cite{DEUMA}.

A {\sl variable} is defined as a set of disjoint attributes. A {\sl task} is represented as a finite set of ordered pairs of input/output attributes: $T=\{{\bf a_1}\rightarrow{\bf b_1}, {\bf a_2} \rightarrow {\bf b_2},\;\cdots\;, {\bf a_n}\rightarrow{\bf b_n}\}$. To illustrate this, let  ${\bf 0}$ denote the attribute for the qubit's state to be in a given subspace and ${\bf 1}$  the attribute for the qubit's state to be in its orthogonal complement. An example of a task is $\{{\bf 0} \rightarrow{\bf 1},{\bf 1} \rightarrow{\bf 0}\}$, negating the qubit in a particular basis. 

Constructor-theoretic statements focus solely on the feasibility or impossibility of tasks, rather than on specific constructors, making them independent of scale and dynamics.

A {\sl constructor} for a task $T$ is a system that, when presented with the substrate of $T$ in a state in one of the input attributes, can transform it into a state within one of the allowed output attributes. Additionally, the constructor {\sl must retain the ability to perform this transformation again}. In chemistry and resource theory, \cite{VIOPLE, CO}, the concept of a catalyst can be seen as a model for specific types of constructors — catalysts must remain in the same state, as opposed to the same attribute, and their behaviour is dependent on dynamics. In quantum theory (see the appendix and \cite{VIOMA}), a constructor is modelled by a subspace $C$ with the following property: the substrate undergoes the transformation specified by $T$, whenever it is coupled to the environment in a state belonging to $C$, and $C$ is invariant under the action of the overall unitary evolution of the joint system of substrates and environment. 

A task is considered {\sl impossible} if the laws of physics impose a limit on how accurately it can be performed by a constructor. If there are no such limits, the task is {\sl possible}. {In quantum information, the physical systems implementing gates serve as examples of constructors: logically reversible computational tasks are all possible under the unitary quantum model of computation. }

{Tasks satisfy algebraic constraints which are detailed in \cite{THE}. In short,} two tasks, $T_1$ and $T_2$, can be composed in series (if the output attributes of $T_1$ match the input attributes of $T_2$) or in parallel, following the conventional meanings of these terms, \cite{DEUMA}. Serial composition is denoted as $T_1T_2$, while parallel composition is written as $T_1\otimes T_2$. The transpose of a task $T$, denoted as $T^{\sim}$, reverses its input/output pairs. It follows that $(T^{\sim})^{\sim}=T$ and $(T_1\otimes T_2)^{\sim}=T_1^{\sim}\otimes T_2^{\sim}$.

It is a principle of constructor theory (one of its proposed laws of physics) that the composition of two or more possible tasks is another possible task. This {\sl composition law} constrains both dynamical laws (for instance, it denies that they limit the depth to which computers could enact recursion) and cosmological initial conditions (for instance, they must provide raw materials that could be assembled into constructors). 

\section{Testing the constructor theory of information}
In this section, we first summarise a few notions from the constructor theory of information and then discuss how it can be tested via the theory of hybrid systems.

The constructor theory of information provides newly conjectured laws for physical systems that can embody information, which are characterised precisely in constructor-theoretic terms by imposing constraints on which tasks must be possible on them. This theory also expresses a physical foundation for the concepts of `distinguishability' and of `measurement', which are usually taken as primitive in the traditional theories of information, classical and quantum. In constructor theory, these concepts are expressed independently of particular dynamical formalisms, only using the set-theoretic structure of constructor theory.

\begin{figure}[h]
	\centering
	\includegraphics[scale=0.5]{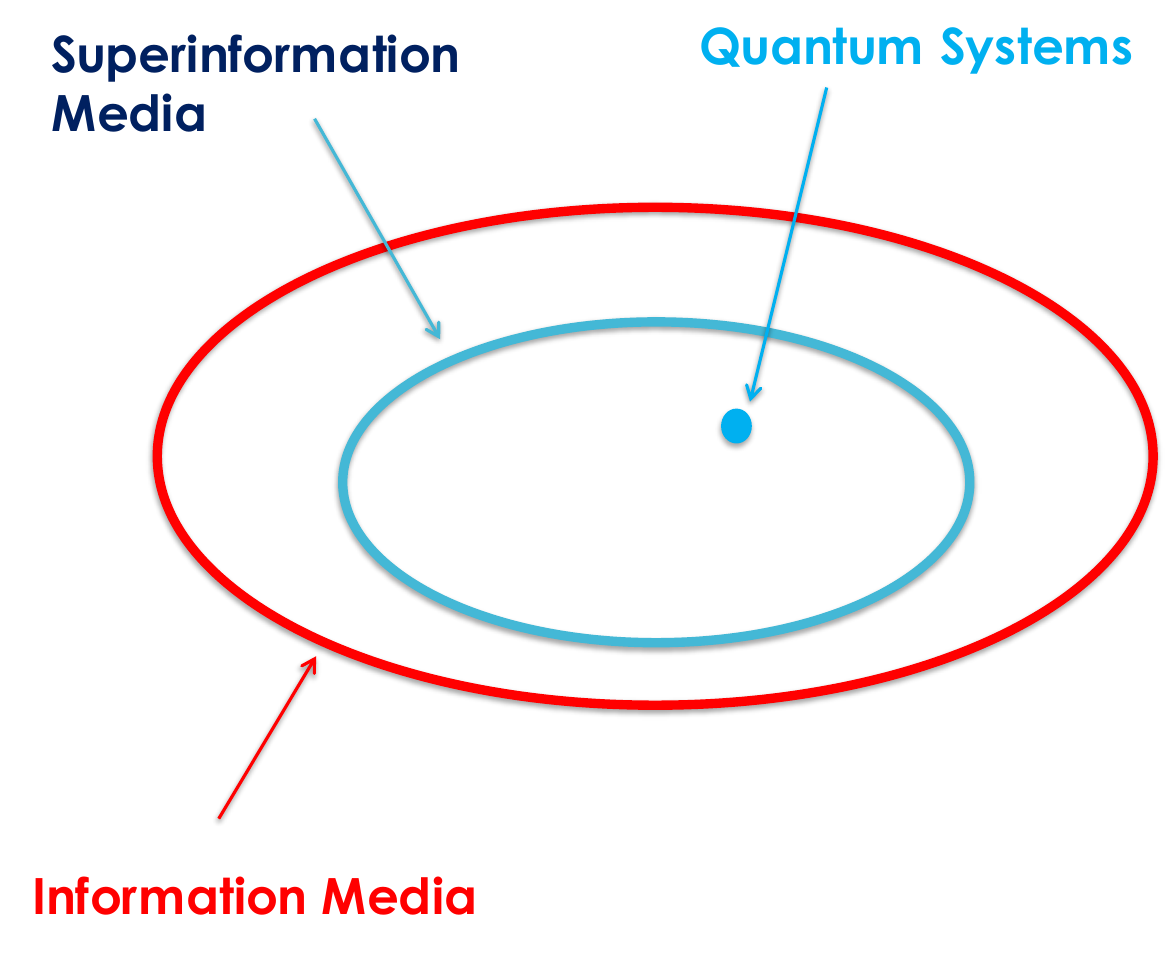} 
	\caption{{Information media and superinformation media: quantum systems can be proven to be a special case of superinformation media. }}
\end{figure}

The dynamics-independent definition of distinguishability, \cite{DEUMA}, is a cornerstone of the constructor theory of information and the basis of the whole construction that defines information media. It generalises the quantum-information notion of states that can be distinguished arbitrarily well from each other with a single-shot, projective measurement (without referring to orthogonality). \footnote{{Other characterisations of distinguishability have since been proposed, see e.g. \cite{REF31}, which assumes the notion of gates to be primitive. Likewise, in \cite{REF32}, the concept of measurement is taken as primitive. Constructor Theory instead grounds both concepts in possible tasks, thus avoiding circularities.}}

First one defines a class of substrates, {\sl information media}, by requiring that some tasks are possible on them - tasks that are conjectured to be sufficient for them to be capable of carrying information. In short, information media must have a variable $X$ with the property that it is possible to perform all the permutation tasks on X, and that it is possible to perform the task of copying all attributes in $X$ from one substrate to its replica (see Appendix A for the formal definitions).  Any variable $X$ that can be copied and permuted in all possible ways is called an {\sl information variable}. {We can consider a qubit to be an example: a qubit is an information medium and any set of two orthogonal states of the qubit is an information variable.}

\noindent A key principle of constructor theory is the {\sl interoperability principle}, which elegantly captures the intuitive notion that classical information should be transferable between any two media of equal capacity, regardless of their physical nature. Specifically, given two information media $S_1$ and $S_2$, respectively with information variables $X_1$ and $X_2$, their composite system $S_1 \oplus S_2$ is an information medium with information variable $X_1\times X_2$, where $\times$ denotes the Cartesian product of sets. 

One can define distinguishability using information media, as follows. A variable $Y$ is {\sl distinguishable} if (informally) it is physically possible to map it onto an information variable in a logically reversible fashion, i.e. if the task

\begin{equation}
\label{eq3}
\bigcup_{{y}\in Y}\left\{{\bf y}\rightarrow {\bf q_y} \right\}
\end{equation}

\noindent is possible, where the variable $\{{\bf q_y}\}$, of the same cardinality as $Y$, is an information variable. Hence, a set of orthogonal quantum states on which the above task is possible constitutes a distinguishable variable. However, we have expressed this concept without relying on the specific formalism of quantum theory, also ensuring that it remains independent of scale and dynamics. This definition therefore is scale- and dynamics-independent, which makes it highly suitable for the kind of tests and systems we shall discuss in the next section.

In addition, in the constructor theory of information, it is possible to express counterfactual constraints on a 
 class of physical systems (called `superinformation media') that can then be proven to exhibit all the typical 
properties of quantum systems. {In particular, the definition of superinformation media says 
that a superinformation medium is an information medium with at least two variables, whose union is not an 
information variable. From this definition, one can derive all the key properties of quantum systems, including: 
the irreducible perturbation caused by measurement, the impossibility of cloning, and the possibility of 
displaying a property that generalises entanglement, called `locally inaccessible information', \cite{DEUMA}. 
Hence, quantum systems are a special case of information media, and superinformation media possess all the 
relevant information-theoretic properties of quantum systems. Examples of superinformation media that are 
not fully fledged quantum systems are, e.g. quantum hybrid models with superselection rules \cite{SUD} and 
non-commuting qubits, \cite{SAM}.} Once more, given that this theory does not rely on the formalism of 
particular dynamical laws, it is an ideal tool to apply in situations where quantum theory's laws may not apply, 
and yet one may have quantum-like behaviours. An ideal context in which to test the constructor theory of 
information is therefore the quantum gravity scenario, or also the coupling between a quantum system and 
another macroscopic system which may collapse away from quantum theory's laws (see the discussion about 
hybrid systems, subsection 2.1). {Remarkably, constructor information theory is scale-
independent, non-circular, and avoids probabilities. This is because the concept of distinguishability is not 
taken as primitive, but is defined in terms of copiability, which is a genuinely physical, non-probabilistic, and 
scale-independent property.}

\subsection{Hybrid systems}

Quantum theory can in principle be applied to objects of any scale and has so far been confirmed experimentally on all fronts. However, some expect it to break down at the macroscopic scale: quantum effects may impossible beyond a given threshold \cite{TH1, TH2, TH3}. This idea forces us to consider hybrid systems, where one subsystem is fully quantum whereas the rest may not be. It is exciting that quantum metrology and quantum information technologies can now investigate hybrid systems in the laboratory. {Typically, experiments involve a quantum probe interacting with a system whose dynamical laws and scale are imperfectly known or intractable. For instance, a large mass in a superposition of locations, interacting with the gravitational field; or a quantum field (e.g., photons) coupled to a macroscopic object whose quantum evolution need not be unitary.}

\begin{figure}[h]
	\centering
	\includegraphics[scale=0.1]{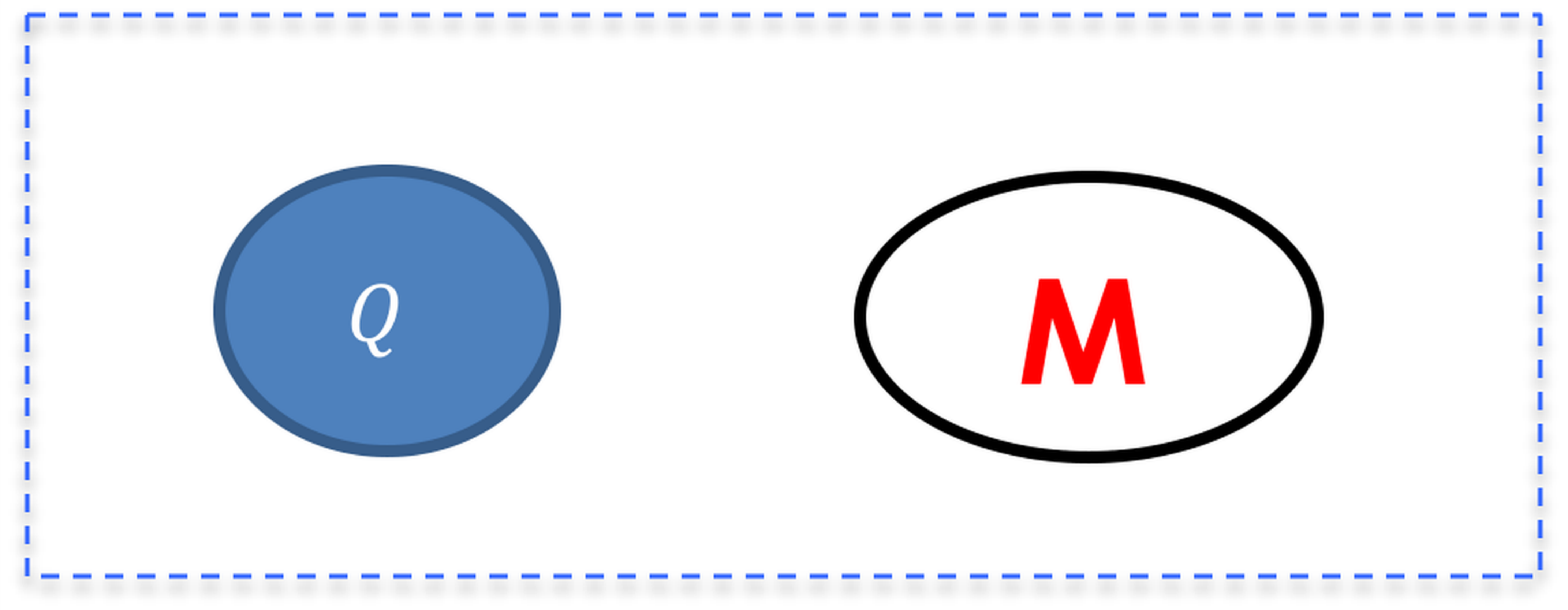} 
	\caption{{Hybrid systems comprise a quantum probe $Q$ (e.g. a qubit) and another `mystery' system $M$ whose laws are partially unknown and may not be quantum-mechanical.}}
\end{figure}

These hybrid systems are characterised by a hybrid scale (not micro or macro, but a mixture of those) and by unspecified hybrid dynamics (not classical, not quantum, but possibly more general ones). Typically, hybrid systems are studied using ad hoc assumptions about dynamics and scale, e.g. via open quantum system dynamics assuming weak coupling, coarse-graining procedures, or semi-classical dynamics. This has the disadvantage of fixing the dynamics a priori. But we know that predictions can also be made via general principles, that are independent of the scale and of the dynamics. Known examples of this principle-based logic are H. Bondi’s derivation of gravitational redshift from the equivalence of mass and energy, \cite{BON}; B. DeWitt’s and R. Feynman’s arguments for quantising gravity, \cite{DEW, FEY}; Einstein’s various derivations in the theory of relativity. Constructor theory generalises this approach and its novel physical principles are ideally suited for hybrid systems. In particular, the theory of superinformation media can be used to treat quantum and classical systems in a unified framework. In the superinformation framework, C. Marletto and V. Vedral have also recast DeWitt's and Feynman's principle-based arguments for why gravity must be quantised in far more general terms, \cite{MAVE17c, MAVE17d}. On this ground, they have recently proposed general witnesses of non-classicality in hybrid systems. These witnesses are the basis of novel experimental proposals, notably C. Marletto and V. Vedral’s proposed test of quantum effects in gravity, \cite{MAVE17a, MAVE17b, MAVEREV}. In the next sections, we shall review these witnesses and the related tests. 

\subsection{Generalisation of DeWitt's totalitarian property}

An interesting question in the context of hybrid systems is whether there can exist a sector of the universe being fully classical, while the rest of the universe obeys quantum theory, while the two can interact in some plausible form. This issue is of crucial importance because the scenario where a fully classical measuring device interacts with a microscopic quantum system may be ruled out if such a general situation were to be impossible, thus showing that quantum theory not only can, but {\sl must} be applicable to measuring devices too. It is well-known that quantum theory can be applied to the measuring apparatus so that everything is consistent, \cite{vonQ, WAL}; but here one would like to find out if this is the only available option, assuming some general principles. 
B. DeWitt proved that it is impossible to have two sectors, one quantum, the other classical, interacting with each other via a measurement-like interaction, \cite{DEW}. This was informally called by DeWitt the `totalitarian property' of quantum theory. In other words, if one part of the universe is quantum, then the rest of the universe must be too, if it is to interact with the former so as to observe it. 

The argument has the weakness of assuming dynamics-specific axioms, including the assumption that the dynamics of both the classical and quantum sectors must be described within the Lagrangian formalism. 

To remedy that, C. Marletto and V. Vedral, \cite{MAVE17c}, have used the constructor theory of information to generalise this result and prove a more general statement: that if a superinformation medium $Q$ interacts with an information medium $M$ by means of a copy-like interaction, then $M$ must also be a superinformation medium. Note that this is more than merely showing that the classical sector must exhibit probabilistic behaviour: superinformation media cannot be emulated by simply stochastic systems, as argued in \cite{DEUMA}. The assumptions of the Marletto-Vedral theorem are the principle of locality, the principle of the interoperability of information (see section 2), and determinism. {The main proof works by contradiction: it is shown that if $M$ were not a superinformation medium, $Q$'s no-cloning principle would be violated as a result (assuming the other principles hold true).}.
Building on this result, C. Marletto and V. Vedral then discovered a method to test for non-classical properties in a physical system by measuring only its classical variables while coupling it to a qubit, \cite{MAVE17d}. Specifically, the experiment involves a composite system consisting of a qubit $Q$ and a classical system $M$, which is assumed to have only a single observable $T$. The proposed test for non-classicality relies solely on measuring correlations involving the observable $T$, without requiring $M$ to undergo interference. {The mediator (M) is called “classical” in a very specific information-theoretic sense, not in the broader sense of ordinary classical mechanics. The key assumption is that (M) possesses only a single observable whose sharp states are all mutually distinguishable. In the simplified qubit-style models used in the theorem, this means that every accessible property of (M) can be expressed as a function of one commuting variable (T). Its state space is therefore effectively that of a classical information carrier — analogous to a classical bit, with states such as ({0,1}). The important point is that “classical” here does not mean “any system described by classical Hamiltonian mechanics.” Ordinary classical systems can certainly have many observables, such as position, momentum, energy, angular momentum, etc. However, all of those observables are jointly sharp and mutually commuting: they can in principle all be simultaneously measured to arbitrarily high accuracy. In that sense, they are all reducible to a single underlying classical observable in phase space.
In constructor theory a system becomes “non-classical” when it possesses at least two observables that are not jointly sharp — the analogue of non-commuting observables in quantum theory. The witness theorem only requires showing that (M) cannot be modelled as a mediator with one purely classical information variable. If (M) can mediate entanglement locally between two quantum probes, then (M) must support additional incompatible observables beyond that single classical variable. This important result led to the next step: proposing the entanglement-based witnesses of non-classicality, which are applicable to testing quantum effects in gravity and other macroscopic systems.}

\subsection{Witnesses of non-classicality in hybrid quantum systems}

The interesting results discussed in the previous section led to the discovery of a witness of non-classicality for hybrid systems and, in particular, to a proposed test of quantum-gravity effects based on constructor theory. 
Instead of referring to ``quantum effects," it is more precise to describe gravity's non-classicality based on the constructor-information-theoretic definition in \cite{MAVE20}. {This is because quantum effects are rooted in the current formalism of quantum theory; while `non-classicality' is a more general property, based on dynamics-independent notions only.}  A system is non-classical if it requires at least two distinct physical variables to describe its behaviour, yet these variables cannot be measured with arbitrarily high accuracy simultaneously. In quantum mechanics, this corresponds to incompatible observables, as described by Heisenberg's uncertainty principle. However, the concept of non-classicality extends beyond quantum theory. It can be precisely defined within the constructor theory of information, which (see section 2.1) generalises quantum information theory to cases where quantum mechanics may not apply. The term ``witness" belongs to the quantum information theory terminology, denoting a Hermitian operator used to determine whether a quantum system is entangled with another \cite{HORHOR}. In general, detecting a witness is a sufficient but not necessary condition for entanglement -- meaning some entangled systems may not satisfy the witness criterion. \footnote{{For instance, a witness on qubit 1 and 2 is represented by the observable $O=X_1Z_2+Z_1X_2$, where $X$ and $Z$ denote the Pauli operators associated with, respectively, components x and z of each qubit. A state is entangled if $\langle O\rangle \leq 1$; the expected value of this observable in the singlet state $\frac{1}{\sqrt{2}}(\ket{01}-\ket{10})$ is zero; thus the singlet fails the witness, despite being maximally entangled.}} Here, we use the term in a similar way: successfully implementing the witness of non-classicality protocol would provide strong evidence that a system is non-classical. However, failing to do so does not immediately prove that it is classical.

This particular witness is based on a general theorem the `General Witness Theorem', (GWT), that can be proven with minimal assumptions about gravity's dynamics, \cite{MAVE20, MAVEREV}, within constructor theory.  We can summarise the theorem this way: if an information medium  ${\bf M}$ (e.g. gravity) can entangle two quantum systems ${\bf Q_A}$ and ${\bf Q_B}$ (e.g. two masses) by local interactions (by ${\bf Q_A}$ coupling to ${\bf M}$ and ${\bf M}$ to ${\bf Q_B}$, but without ${\bf Q_A}$ and ${\bf Q_B}$ interacting directly) then ${\bf M}$ must be non-classical. The theorem, in this general form, is a generalisation of the Local-Operation-and-Classical-Communication (LOCC) theorems of quantum information, \cite{HORHOR} (which assume quantum theory's formalism) to the domain where quantum theory may not apply. {The central setup is that two quantum systems, $(Q_A)$ and $(Q_B)$, interact only locally with a mediator (M), with no direct interaction between the probes themselves. The probes are assumed to be genuinely quantum, in the sense that they possess complementary observables and can therefore become entangled. In contrast, it is not assumed that (M) is described by a Hilbert space, obeys quantum dynamics, or even corresponds to a standard classical mechanical system. Instead, the theorem assumes only that systems can exchange and carry information consistently through local interactions, as required by the constructor theory of information. The crucial assumption concerns what it means for (M) to be “classical.” In the GWT, a classical system is one whose observables are all jointly sharp, meaning that all accessible properties can be reduced to a single commuting informational variable. Equivalently, (M) has distinguishable states and probabilistic mixtures over them, but no incompatible observables analogous to non-commuting quantum operators. The theorem then shows that if entanglement is nevertheless generated between $(Q_A)$ and $(Q_B)$ through local interactions with (M), this kind of purely classical mediator is impossible. Therefore, (M) must possess at least two incompatible observables, or some equivalent form of non-classicality. Importantly, the conclusion is not that (M) must already satisfy the full formalism of quantum theory, but only that it cannot be entirely classical in the operational sense assumed by the theorem.}{The logic of the proof is to show that if the two probes can get entangled locally via $M$, then $M$ cannot be purely classical, otherwise this would violate the no-cloning principle on the two probes.}.
\begin{figure}[h]
	\centering
	\includegraphics[scale=0.1]{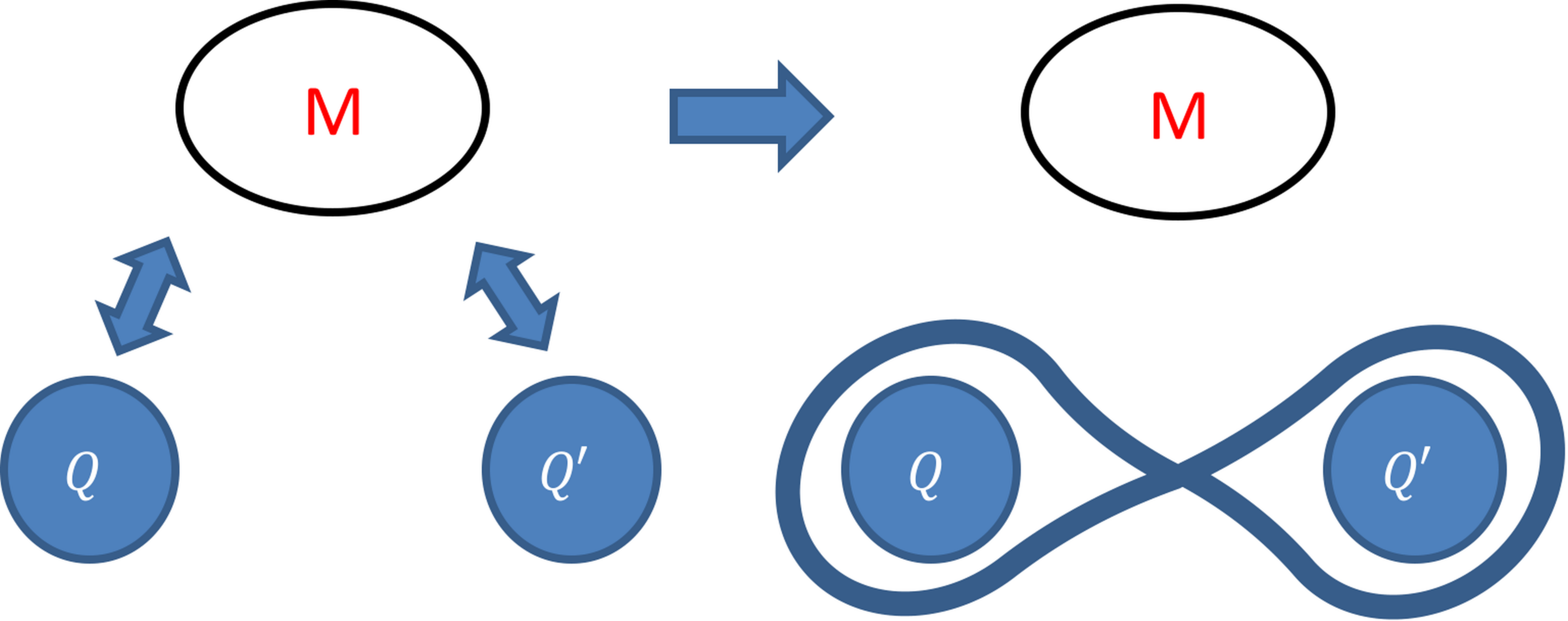} 
	\caption{{The entanglement-based witness of non-classicality. If M can entangle (by local means) the quantum probes Q and Q', then it must be non-classical.}}
\end{figure}

The GWT rests on two very general principles. One is the principle of locality, or {\sl no action at a distance}, that has to be satisfied by the dynamical theories describing both the entanglement mediator and the probes. This principle, which can be expressed in constructor theory (see \cite{DEUMA, THE}, then allows one to assume pairwise interactions of each quantum probe with the mediator, but not directly with each other).  
The other is the principle of {\sl interoperability of information} (see section 2). As we explained earlier, this principle requires that the composite system of two systems that can each contain information must be capable of containing information, too. For instance, two systems each capable of containing one bit of information, when considered jointly, can hold two bits worth of information. 

{Due to its generality, the GWT is akin in spirit to Bell's theorem, \cite{BELL}, which has been a cornerstone of ruling out classical models for quantum statistics. By violating Bell's (and related) inequalities in an experiment, one can rule out a vast class of local hidden-variables, real-valued stochastic models. Likewise, when observing entanglement in the context of the witness, one can rule out for the mediator all classical theories obeying the above-mentioned general principles, \cite{MAVE18}. This includes classical theories that are known (e.g., quantum field theory in curved spacetime, general relativity, collapse models) and even those classical theories that are yet to be discovered.}

\subsubsection{Applications to gravity: the BMV experiment}
 
The BMV (Bose-Marletto-Vedral) experiment, designed to probe quantum aspects of gravity, is directly based on the entanglement-based witness of non-classicality. Consequently, it also stems from the Constructor Theory of Information—specifically, the principle of the interoperability of information. As a result, this yet-to-be-performed experiment would serve as an indirect test of the validity of the interoperability principle. 
The fundamental idea is to study gravity as a potential mediator of entanglement, using two quantum systems (probes) (see the review \cite{MAVEREV} and references therein). The original proposal involves two masses, each placed in a superposition of positions. If gravity can entangle these quantum probes (under specific assumptions discussed earlier: locality and interoperability of information), it must exhibit non-classical properties. Informally, this implies that gravity cannot be fully described by a classical theory such as general relativity. This phenomenon is referred to as gravitationally induced entanglement (GIE) \cite{MAVE17b}. A key breakthrough of the GIE proposal is that it suggests signatures of quantum gravity can be detected at scales below the Planck regime. In fact, the GIE experiment is particularly promising because it involves masses significantly smaller than the Planck mass—on the order of nanograms—making it more experimentally feasible than previous proposals {(see \cite{MAVEREV}} for a discussion) {due to the low-energy ensuing requirements.}

This feasibility arises because the experiment examines the quantum nature of gravity indirectly through quantum probes, rather than attempting to detect individual quanta of the gravitational field, such as gravitons, predicted by quantum gravity theories.

A notable quantum simulation of constructor theory’s predictions regarding the non-classical mediation of entanglement via gravity was carried out by J. Jones and his team using NMR qubits \cite{Jones}. {Notably, the simulation proposes a qubit model for the BMV experiment; it also provided a quantitative model of the transition from a full non-classical behaviour (a qubit) to a fully classical system (bit), by using NMR qubits affected by noise.}

The BMV effect is rooted in the GWT, a constructor-theoretic result that lends robustness to the test. However, the generality of the information-theoretic approach comes with a trade-off: GIE cannot be used to falsify specific {dynamical theories of quantum gravity}. In the quasi-Newtonian regime where GIE emerges, existing non-perturbative frameworks such as string theory and loop quantum gravity concur with the predictions of linearised quantum gravity. Consequently, should entanglement be observed, the witness does not distinguish one quantum gravity theory from another.

{Thus constructor theory is not used to formalise a model for the experiment (the only viable one at present is linear quantum gravity, \cite{MAVEREV}).}{Rather, it is used to express the GWT -- an argument to show that the experiment works as a general witness of non-classicality,  to maximise its applicability. Using LOCC arguments (like in \cite{BOS})  or a more general argument still formulated within quantum theory (like in \cite{MAVE17b}) makes the test only narrowly applicable: under such arguments, observing GIE would simply rule out models of gravity where the latter is treated quantum-mechanically and then decohered. With the GWT, instead, the argument becomes broader and applies to all theories that satisfy the axioms (locality and the interoperability of information), not necessarily only to those expressed within quantum theory’s formalism. This is what makes the BMV experiment more robust.}

If GIE were successfully detected, it would provide the first experimental contradiction of Einstein’s general relativity as a purely classical theory of gravity, reinforcing the necessity for modifications incorporating quantum features, as suggested by existing quantum gravity models. It would also, indirectly, corroborate the interoperability principle applied to the composite system of gravity and the two quantum probes. This could have interesting implications for considering quantum-information-processing devices based on gravitational interactions. {The principle of interoperability of information is not the sole assumption behind the BMV experiment or the General Witness Theorem. Rather, it is part of a broader operational framework that also assumes locality of interactions, the existence of composite systems, and the ability of the probe systems to exhibit genuinely quantum behaviour such as complementarity and entanglement. Interoperability plays an important role because it guarantees that information can be consistently transferred and composed between physical systems, allowing one to reason about mediation and entanglement operationally without assuming the full formalism of quantum theory. However, the experiment does not test interoperability alone. Instead, the logic is conditional: if entanglement between the probes is observed and all interactions are local, satisfy the interoperability condition, and are mediated solely by gravity, then the mediator cannot remain purely classical in the information-theoretic sense used in the theorem. Conversely, the absence of GIE in a carefully prepared experiment would raise several equally interesting possibilities. It could indicate that gravity is fundamentally classical, challenging the completeness of quantum theory at certain scales. Alternatively, gravity may indeed be quantum, but our current theoretical frameworks might be insufficient to predict this effect. Another possibility is that the principles of the constructor theory underlying the non-classicality witness are incorrect: this could refute either locality or the interoperability of information. Any of these outcomes would have profound implications for the future of fundamental physics as it would open up unanticipated research programmes to solve each of those problems}.

\subsection{Temporal witness of non-classicality}

It is possible to consider a `temporal' version of the GWT, where the mediator $M$ does not mediate entanglement between two spatially separated systems, rather it mediates the dynamical evolution of a single quantum probe $Q$. This idea leads to the so-called `temporal witnesses of non-classicality', which G. Di Pietra and C. Marletto have investigated in \cite{DIMA1}. The witness theorem underlying this protocol is this: If an information medium $M$ can induce a quantum evolution on a probe $Q$, and a global quantity on $M$ and $Q$ is conserved, then $M$ must be non-classical. {As always, given the reduction in complexity (from two, to one probe), there is a corresponding increase in assumptions. In particular, one needs extra axioms to prove that this protocol is a witness of non-classicality for the mediator.} One extra assumption is the conservation of some quantity (energy, momentum or any other additive quantity). This law here plays the role of the locality principle in the spatial version of the GWT. The other assumption is the formalism of quantum theory -- however, we believe that the latter can be relaxed, delivering a fully constructor-theoretic generalisation of this argument. The applications of this witness could include a macroscopic bio-molecule as $M$, and the photon field as the quantum probe, \cite{DIMA2}. In that case, measuring a quantum-coherent evolution of the quantum probe would lead, under the assumption of the conservation law, to concluding that the mediator of that evolution must be non-classical.

\begin{figure}[h]
	\centering
	\includegraphics[scale=0.1]{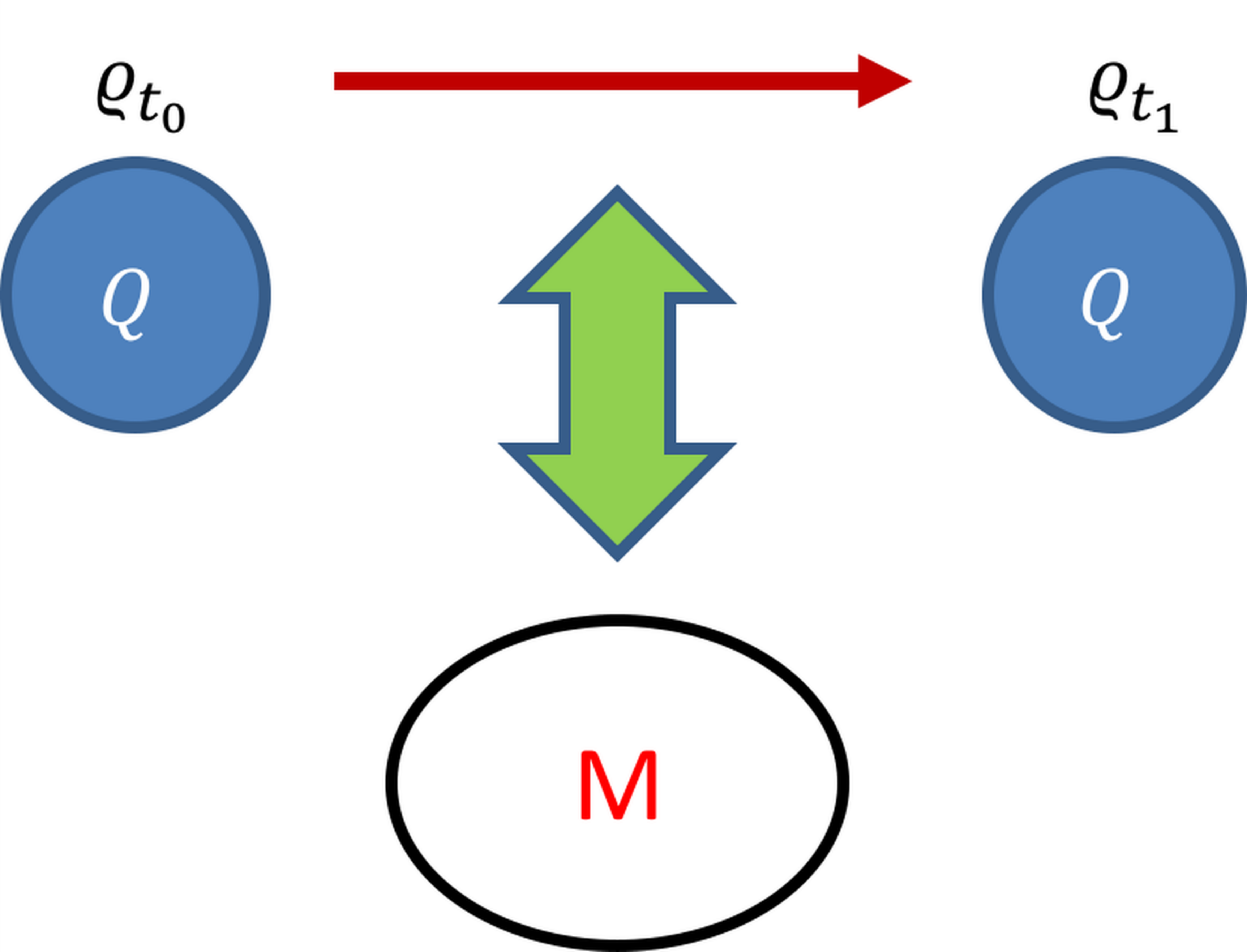} 
	\caption{{ Temporal witness of non-classicality. If $M$ can induce a quantum dynamical evolution on the quantum probe $Q$, then, upon assuming the conservation of an additive quantity, $M$ must be non-classical.}}
\end{figure}
The conservation law here could be that of energy which naturally leads us to the topic of thermodynamics. In the next section, we discuss to phrase the laws of thermodynamics in the most general way, using constructor-theoretic statements, and survey the relevant experimental demonstrations conducted so far to test a novel kind of irreversibility stemming from the constructor-theoretic take on thermodynamics.

\section{Constructor Theory of Thermodynamics}
In this section, we discuss briefly the key results of the constructor theory of thermodynamis, and then discuss its proposed and performed tests so far. 

Thermodynamics constitutes an essential part of the edifice of theoretical physics, yet it is affected by the problem that its laws are, as currently formulated, scale-dependent. Also, deriving or even reconciling the second law with the microscopic dynamical laws is notoriously problematic. Much progress has been made along the lines of deriving the statistical mechanics second law in quantum theory, from certain typicality arguments; and also in generalising classical laws of thermodynamics to quantum dynamics, with the hotly investigated field of quantum thermodynamics, \cite{THERMO, LOST}. Constructor theory provides an alternative route to generalising thermodynamics, so that it becomes scale independent and, crucially, it holds for a vast class of dynamical laws, not just quantum theory. The starting point is the axiomatic approach to thermodynamics, \cite{LIE}. It expresses the second law as requiring that there are tasks $A\rightarrow B$ that can be performed with mechanical means only, to arbitrarily high accuracy, (e.g. heating up some water by stirring only), whose transpose task $B \rightarrow A$ is impossible to perform by such means only (e.g., a heat sink is needed to cool the water to arbitrarily high accuracy). Tasks that are possible to perform with mechanical means only are termed `adiabatically possible'. Still, like all classical formulations of thermodynamics, this axiomatic formulation too only applies at a certain ill-defined `macroscopic’ scale: its applicability to microscopic systems is somewhat controversial, \cite{UFF}, due to the ad-hoc, macroscopic nature of the concepts of adiabatic accessibility and of mechanical means. 

Remarkably, in constructor theory it is possible to express the first and second laws of thermodynamics in the axiomatic approach, expanding on the latter to make it independent of the scale and also of the particular dynamics \cite{MAT1, MAT2}. This can be done by improving on the condition of adiabatic accessibility and mechanical means, by expressing them via the novel construct of `work media', which we shall discuss in section 3.1. It was also recently shown that this constructor-theoretic formulation can be reconciled with time-reversal symmetric laws exactly, with no approximation schemes, \cite{VIOMA}. The reason is that it requires certain tasks to be impossible, not that time-reversed trajectories be forbidden.\footnote{{We note that there are other, statistical approaches to thermodynamics; in such formulations, the second law is not a law of impossibility, but it is one of improbability.}} 
This also establishes an unexpected novel connection between the first law (conservation of energy) and information theory. Building on this work, a model for a novel kind of constructor-theoretic irreversibility in unitary quantum theory has been proposed, where the task of converting the quantum state A into state B is possible, but the transpose task $B\rightarrow A$ need not be, using the quantum homogeniser (a unitary, qubit-based simulator of thermalisation), \cite{HOM}. We will now summarise the main points of these interesting experimental results.

\subsection{Work media}

In traditional thermodynamics, there is a general consensus, following Planck, on identifying mechanical means (or a work repository) with a system behaving `in the same way' as a weight in a uniform gravitational field, which can be smoothly raised or lowered to different heights, \cite{BUC, UFF}. This definition has two problems: 1) it is ad-hoc, appealing to a weight in a gravitational field; 2) it is scale-dependent. In quantum thermodynamics and related resource-theoretic accounts, it is common practice to define a work repository as a system in any eigenstate of its free Hamiltonian, such as a set of bound states in an atom, usable as a battery. The disadvantage of such definitions is that they are rooted in a particular dynamics and formalism (quantum theory's) and thus lose generality compared to the original classical thermodynamics' formulation.  {Other resource-theoretic accounts, where quantum theory is not assumed, e.g. \cite{SCA}, are rooted within other dynamical formalism, so they are, still, dynamics-dependent. There are also other existing approaches -- see e.g. \cite{RREF} -- where irreversibility is built-in at the foundations. Instead, constructor theory allows for irreversibility but does not impose it. Indeed, it is compatible with a universe without any arrow of time.}

To solve the above issues in current approaches, in constructor theory, we can be more general than existing mechanical means/work repository notions, but compatible with all of them. In particular, C. Marletto has demonstrated that it is possible to generalise the class of mechanical means to that of {\sl work media}, \cite{MAT1, MAT2}. As we shall now explain, these are a particular class of substrates satisfying an operational criterion (just like information media): certain tasks must be possible on a substrate for it to qualify as a work medium. This provides a conjectured scale- and dynamics-independent generalisation of the notion of work repository, building on the classical definition of Planck's and Clausius', which improves both on the classical thermodynamics approaches (making them scale-independent) and on the quantum thermodynamics/statistical mechanics ones (making them dynamics-independent). This, in turn, provides the foundations for a scale-independent expression of adiabatic accessibility and thus of the second law of thermodynamics in the axiomatic approach.  

The definition of work media proceeds in two steps. First, one needs to express the conservation of energy in constructor theory. This is done by showing that the presence of a conservation law implies that tasks on a given substrate are partitioned into {\sl equivalence classes}. Here we shall call these classes `energy-equivalence-classes', as we will focus on energy conservation only. Tasks belonging to the same equivalence class violate the energy conservation by the same amount - see \cite{MAT1, MAT2} for the formal details.

Then, one gives sufficient conditions to identify work media, as follows. { A work medium is a substrate ${\bf Q}$ having a variable $W=\{{\bf w_+}, {\bf w_0}\}$ with the property that:}

\begin{itemize}

\item{i)} The task $T_{+,0}=\{{\bf w_+}\rightarrow{\bf w_0}\}$ belongs to an energy-equivalence class such that $T_{+,0}$ is impossible and so is its transpose. 
\item{ii)} There exists an attribute ${\bf w_-}$ of ${\bf Q}$, disjoint from ${\bf w_0}$ and ${\bf w_+}$, such that the task
\begin{equation}
\{({\bf w_+},{\bf w_0}) \rightarrow ({\bf w_0},{\bf w_+}), ({\bf w_0},{\bf w_0}) \rightarrow ({\bf w_+},{\bf w_-})\} \;\;\label{SWAP}
\end{equation}
defined on the composite substrate made of to replicas of {\bf Q} is possible.

\end{itemize}

\noindent The variable $W$ is called a {\sl work variable}.

{The compensating task condition in equation \eqref{SWAP} formalises the intuition that work can be stored and transferred in a controllable way on all work-media, in a scale- and dynamics-independent manner. A work medium, therefore, behaves analogously to an idealised weight in classical thermodynamics: transitions between its attributes can systematically compensate one another across different substrates. This is why it can be used to define adiabatic accessibility in a more general way than in the traditional approaches. Heat media lack this property because their state changes are not fully recoverable through controlled pairwise compensations. In particular, the condition (ii) expresses the existence of a compensating transformation on a replica of {\bf Q}, required so that the overall task is possible in the presence of the conservation of energy. }

{For example, a system possessing a work variable is an atom $Q$ with three different equally-spaced energy levels, in decreasing order of energy as follows: ${\bf w_+}, {\bf w_0}, {\bf w_-}$.} 
Given finite resources, executing the task $T_{+,0}$ is impossible due to energy conservation, as it necessitates a change in the atom's energy. Since energy conservation mandates that any finite-dimensional environment interacting with the atom must experience an energy shift equal and opposite to that induced by $T_{+,0}$, the task is impossible on {\bf Q} only. Thus, condition (i) holds. Moreover, the task in \eqref{SWAP} can be realised through an appropriately designed energy-preserving unitary transformation. Therefore, a quantum system with at least three equally spaced energy levels satisfies the criteria for work media, aligning this definition with well-established classical and quantum notions of a work repository.

The work media definition precisely captures the characteristics required to extract energy from or transfer energy to another system {\sl reversibly}, without introducing any additional side effects. It extends the notion of `work repository' or`mechanical means' to general systems beyond mechanical ones, such as an atom in an excited state. This definition refines existing approaches, which often designate energy eigenstates as work repositories by assumption. As a result, it is the base for a scale- and dynamics-independent framework for deterministic work extraction.  {It also provides the new definition of adiabatic accessibility, as mentioned earlier, which allows one to express the second law in a scale- and dynamics- independent way (see \cite{MAT1, MAT2} for details). }

{Thus, in constructor theory, work and heat are not distinguished primarily by “energy transfer,” but by the  kind of tasks that are allowed on a physical substrate. The defining feature of a work medium is not simply that transitions between attributes involve different energies, but that the change of energy is accomplished by a side-effect on a replica of the same medium, according to the condition (ii) as expressed in equation (3). Crucially, condition (ii) is {\sl not} satisfied by purely thermal attributes such as having a particular temperature, in line with traditional thermodynamics: in other words, a single thermal state cannot do work.  For example, let's assume ${\bf w_{\alpha}}={\bf T_{\alpha}}$, where the attributes ${\bf T_+}, {\bf T_-}, {\bf T_0}$ of, say, a volume of water each correspond to a thermal state with given temperature $T_{\alpha}$. In order to satisfy the first requirement (equation \eqref{SWAP}),  an equilibrium state $({\bf T_0},{\bf T_0})$ should be transformed into the temperature attribute $({\bf T_+},{\bf T_-})$, with no other side effects. This is impossible according to the second law in classical and quantum thermodynamics. Thus, systems endowed with thermal degrees of freedom within the standard definitions of thermodynamics do not qualify as work media.}

\subsection{Testing irreversibility in a time-reversal symmetric universe}

Using the powerful apparatus of constructor theory, one can conjecture the following ``constructor-based" irreversibility, \cite{VIOMA}. Consider a possible task $T$, requiring that some input attribute $\textbf{x}$ is modified into an output attribute $\textbf{y}$. Its transpose $T^\sim $ is defined to have the inputs and outputs of $T$ switched, so it has input attribute $\textbf{y}$ and output attribute $\textbf{x}$. The constructor-based irreversibility consists in the fact that if the task $T$ is possible, its transpose $T^\sim $ need not be possible, even with time-reversal symmetric dynamics \cite{VIOMA}. 

This is because the existence of machines that can approach a perfect constructor for a task $T$ {\sl does not imply} that such machines exist also for the task $T^{\sim}$ in the reverse direction. For machines approximately capable of performing a task in a cycle are not necessarily able to perform the transpose task in a cycle to the same degree of approximation simply by having their dynamics reversed. This asymmetry exists even if the underlying dynamical laws are time-reversal symmetric, like unitary quantum theory's, (as proved in \cite{VIOMA}).

This result led to an interesting experimental demonstration of the constructor-based irreversibility. In particular, a model was recently proposed to illustrate the constructor-based  irreversibility, based on unitary quantum theory \cite{VIOMA, MAGE}. Specifically, in this model, it is demonstrated that transforming a qubit from a pure state to a mixed state is possible, while transforming a qubit from a mixed to a pure state is not necessarily possible, even if this transformation is done via a series of unitary, time-reversal symmetric interactions. The model is based on the quantum homogeniser, which is a quantum computing model originally proposed to model the processes of erasure and thermalisation in a reversible fashion, using unitary gates called `partial swap', \cite{HOM}. 

In \cite{VIOMA} was also shown that in the weak coupling limit, the quantum homogeniser can only be used as a constructor for the pure-to-mixed task and not the mixed-to-pure task. To assess whether a task can be performed with arbitrary accuracy in a cycle by homogeniser, a novel measure of irreversibility was introduced, called {relative deterioration} \cite{VIOMA, MAGE}. This quantity is a function of both the { error} in performing the task, to quantify the potential for arbitrary accuracy, and of the { robustness} of the machine with multiple iterations, to quantify the potential for performing the task again once it has been performed once.  

In collaboration with C. Marletto and V. Vedral, M. Genovese and his experimental team also performed an experiment with photonic qubits, demonstrating the compatibility of such irreversibility with quantum theory’s time-reversal symmetric laws, \cite{MAGE}. The experiment adopted the dynamical model based on the universal quantum homogeniser (see the picture) to show the dynamical evolution of the relative deterioration. They then tested the physical realizability of this model by means of an experimental demonstration with high-quality single-photon qubits. We note that {the asymptotic limit appearing in the homogeniser model is not in the expression of the constructor-theoretic irreversibility (the latter is scale-independent: it has no limits in it). Rather, it pertains to the particular model we used to instantiate a constructor for a given task (the homogeniser). What is scale-independent is the statement that there exist tasks that are (adiabatically) impossible in one direction, but (adiabatically) possible in the reverse (this is a scale-independent expression of the second law). The homogeniser model we studied is an instance of this irreversibility. In the model, the perfect-constructor behaviour for the task in one direction is attained in the infinite limit. The irreversibility, however, is present at every finite step (by continuity), which is what the simulation with a finite number of qubits has experimentally demonstrated.}
\begin{figure}[h]
	\centering
	\includegraphics[scale=0.5]{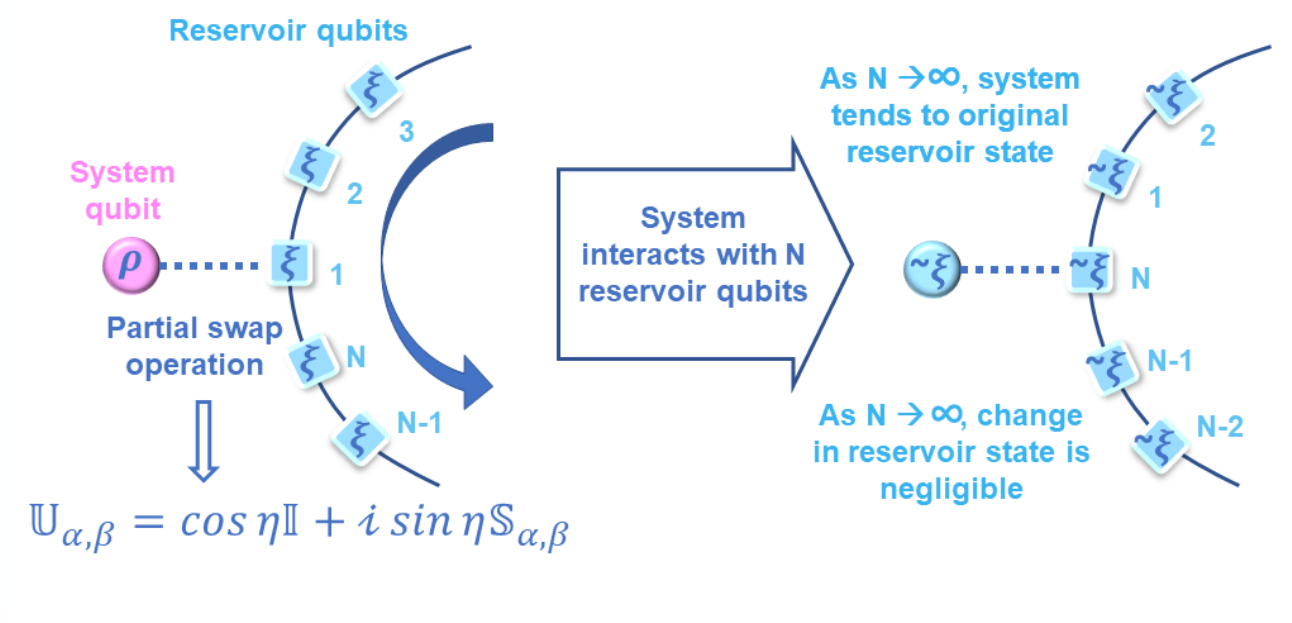} 
	\caption{The quantum homogeniser used to model the constructor-based irreversibility. {The homogenizer consists of $N$ qubits initialised in the target state, $\ket{\xi}$. The system qubit is initialised in the input state $\ket{\rho}$. At each step, the system qubit interacts with one of the qubits in the homogeniser, via the partial swap operation (parametrised by the angle $\eta$):  $\mathds{1}$ is the identity transformation, while $S_{\alpha, \beta}$ is the swap gate defined on two qubits $\alpha$ and $\beta$. As $N\rightarrow \infty$, the system qubit approaches the target state after $N$ partial swaps.}}
\end{figure}

This experiment has therefore shown that the constructor-theory irreversibility is compatible with quantum theory's reversible, unitary dynamical laws. This result can be used as the basis for future experiments to investigate, e.g., how this irreversibility differs from the standard, entropy-based irreversibility in quantum and classical thermodynamics. 

\section{Other testable aspects of constructor theory}

{This section will identify the areas of the theory where additional tests could be designed. These tests will only be outlined at the most general level, leaving detailed experimental proposals for further work.}

\subsection{Testing the Born Rule}
In constructor theory, C. Marletto has shown under what conditions theories describing superinformation media can provide the same probabilistic structure as quantum theory, \cite{MAP}: this work comprises an alternative derivation of the Born Rule that replaces decision-theoretic axioms proposed by D. Deutsch and then D. Wallace, \cite{DEUPRO, WAL}, with some of the information-theoretic principles of constructor theory. {The Born Rule in constructor theory is derived from non-probabilistic constraints about possible/impossible tasks, not assuming particular dynamics or state-space structures, in sharp contrast with other derivations, such as \cite{PROB1, PROB2}.} The constructor theory of probability provides new principles constraining superinformation theories -- i.e., theories that describe superinformation media. Those that satisfy these constraints have the potential to generalise quantum theory as they exhibit the same Born-Rule-like appearance of stochasticity. Superinformation theories allow for all the disparate properties of quantum information (including entanglement). Unlike existing stochastic generalisations of quantum theory, e.g. Barrett’s generalised probabilistic theories, \cite{BAR}, superinformation theories are local (in the strong sense of Einstein’s locality criterion, as satisfied by quantum field theory) and deterministic: probabilities are only emergent, approximate properties. Consequently, superinformation theories capture exactly the information-theoretic features of quantum theory and are ideal candidates to describe quantum hybrid systems, as defined in section 2.1. {The axioms that Marletto uses can be summarised as follows. First, there must be observables that are incompatible, as they cannot be copied simultaneously to the same arbitrarily high accuracy, as in superinformation media.
Marletto then introduces the idea of X-indistinguishability classes, where $X$ is a given observable. When one considers measurements of an observable $X$ on sets of identically prepared systems, different microscopic states must become operationally indistinguishable in the asymptotic limit via the frequencies of outcomes they generate under repeated measurements. These equivalence classes behave mathematically like probability distributions, even though no primitive probabilities have been postulated. Probability emerges as a higher-level description of what observers can consistently infer from repeated experiments.
The framework also requires certain symmetry properties. Different measurement arrangements must be physically equivalent and thus considered the same within an experiment. These symmetries underpin the assignment of equal weights in situations analogous to equal-amplitude quantum superpositions.
Interestingly, such conditions are purely information-theoretic and are not decision-theoretic, as opposed to previous approaches.}

This work on the Born Rule in quantum theory showed that superinformation theories (and thus unitary quantum theory) exhibit unpredictability in certain measurements yet remain deterministic. The unpredictability {(which is due to the impossibility of certain tasks, and is compatible with determinism)} is thus precisely distinguished from randomness, and it is linked directly with the impossibility of copying certain sets of states. C. Marletto's derivation of the Born Rule also improves on previous decision-theoretic approaches in the following two ways: 1) Its axioms, formulated in constructor-information-theoretic terms {only}, make no use of concepts specific to Everettian quantum theory, thus broadening the domain of applicability of the approach to decision-supporting superinformation theories; 2) It shows that some of the assumptions that were previously considered as purely decision-theoretic, and thus criticised for being ‘subjective’, follow from the {physical properties }of {superinformation media}, {measurers }and {adders}, as required by the newly proposed laws of constructor theory.
This result, together with D. Deutsch's recent paper on testability, \cite{DEUTEST}, implies that it is possible to regard the set of decision-supporting superinformation theories as a set of theoretical possibilities for a local, non-probabilistic generalisation of quantum theory, thus providing a new framework where the successor of quantum theory may be sought.
{In the context of this review, this work is highly relevant because it allows one to design alternative tests of the Born Rule. In particular, one possible line of testing is as follows. First, one can identify the superinformation theory probability axioms, which can be deformed so that the modification can be quantified by a real-valued number. Once this parametrisation is in place, one can run the argument in  \cite{MAP} and obtain a modification of the Born Rule parametrised by that number. The non-trivial step is to identify deformations that preserve the internal consistency of superinformation theory while inducing controlled deviations in the emergent probability rule. Once the parametrisation is established and a modified Born Rule derived, it is important to clarify what confirming the Born Rule could imply for the axioms of constructor theory. One could add a theoretical argument to explain why such axioms, with additional assumptions, are necessary (and not just sufficient) for the Born Rule to ensue. This would close the loophole and allow one to claim that testing the Born Rule indirectly provides evidence for the constructor-theoretic axioms.}

\subsection{Constructor Theory of Life}

John von Neumann, in his groundbreaking work on universal constructors and self-replicating machines, identified the fundamental `replicator–vehicle' logic underlying self-reproduction (see Figure 5). {However, due to the prevailing perspective at the time, he was compelled to frame his analysis in predictive terms, using transition rules in some dynamical model, instead of adopting counterfactuals}. This led to an unsuccessful attempt at designing a true self-replicating system using atoms and microscopic interactions. Ultimately, he devised a functional toy model within cellular automata, but this abstraction severed its connection to real-world physics. As a result, the model does not resolve the core question of whether self-reproduction can emerge under actual physical laws without relying on pre-engineered adaptations.

In the constructor theory of life, \cite{MAL}, C. Marletto showed that under no-design laws,  a high-fidelity replicator is necessary, not only sufficient, for accurate self-reproduction of complex objects. {The argument begins with the problem of how accurate self-reproduction of a complex entity could occur under what she calls “no-design laws” — laws of physics that do not already contain the design of biological adaptations encoded within them. The next move is to analyse reproduction in constructor-theoretic terms, as a network of tasks. A self-reproducer is decomposed into two logically distinct components: a replicator, which stores the information specifying the organism, and a vehicle, which performs the physical operations needed to build and maintain the organism and copy the replicator (see Figure 6). This mirrors the biological distinction between genes and cellular machinery. The argument then shows that accurate self-reproduction cannot be achieved by “template copying” of the whole organism alone. If every feature of the offspring had to arise directly from the dynamical laws or from detailed copying of the parent’s entire structure, then the laws themselves would effectively have to encode the organism’s design. That would violate the no-design condition. Instead, the design information must exist in a modular information medium that can be copied independently of understanding or simulating its content. This informational component is the replicator. It carries abstract instructions that can be copied with high fidelity because the copying process depends only on the physical form of the information-bearing states, not on the semantic complexity of what they encode. The vehicle then executes those instructions and constructs the organism. Crucially, the vehicle's architecture itself is also specified by the replicator, so the system closes recursively. This result explains why digital genetic information is not merely contingent biology but follows from general physical principles. Replicators emerge as the necessary way via which accurate reproduction of complex entities and cumulative evolution occur without hiding biological design in the laws themselves.
} This result confirms that von Neumann's replicator–vehicle logic {is not only sufficient, but also necessary, for accurate self-reproduction under such laws}. Unlike other approaches, such as Assembly Theory \cite{SWLC} (which aims at quantifying the complexity of given objects considering how hard it is to assemble them from scratch), constructor theory aims at defining fundamental constraints under which life can emerge under physical laws, including ours, rather than at predicting the evolution of the biosphere on this planet. However, the principles of the constructor theory could be used, much like the second law of thermodynamics, to impose constraints on how evolution may pan out, given that it must satisfy the principles. For example, one could make predictions about how likely it is for a given form of life to continue existing, once it has come into being, considering its given environment. Crucial to this kind of prediction is the theory of knowledge, which is currently under development in constructor theory, \cite{DEU25}. {Possible experiments to test the constructor theory of life are as follows. First, one could attempt to quantify the relationship between knowledge in the replicator (defined as information that can act as a constructor, \cite{DEU}), the relative deterioration of an organism or system in a given environment (as defined in \cite{VIOMA}), and the level of environmental noise present in that environment. The central conjecture is that systems exhibiting lower relative deterioration under noisy or fluctuating conditions must necessarily contain a higher degree of knowledge. (Within constructor theory, knowledge is understood not merely as stored information, but as information capable of causing (or preventing) transformations, including maintaining its own persistence under laws of physics that are not specifically designed to preserve it.)
One possible experimental approach would involve comparing biological or artificial self-maintaining systems subjected to controlled environmental perturbations. Such perturbations could include thermal fluctuations, chemical instability, radiation exposure, or informational noise affecting signalling pathways. The rate of deterioration could then be measured through quantities such as replication fidelity, structural stability, metabolic efficiency, or persistence time. Simultaneously, one would require a quantitative proxy for constructor-theoretic knowledge content. This might be approximated through measures of algorithmic complexity, error-correction capacity, adaptive flexibility, or the amount of information required to reproduce the system’s maintenance processes. The conjecture predicts a systematic relationship between these quantities: systems capable of maintaining themselves in increasingly noisy environments should require increasingly sophisticated knowledge-bearing structures. For example, organisms with more elaborate repair mechanisms, regulatory networks, or adaptive behaviours may exhibit lower relative deterioration despite higher environmental noise. Conversely, systems with limited knowledge content should deteriorate more rapidly once environmental perturbations exceed a certain threshold. Such experiments would not directly verify constructor theory in its entirety, but they could test one of its central explanatory claims: namely, that the persistence of life is fundamentally linked to the existence of knowledge embodied in physical substrates capable of acting as constructors, under no-design laws. If a robust quantitative relationship between knowledge and resistance to deterioration were observed across different systems and environments, this would provide empirical support for the constructor-theoretic account of life as an emergent phenomenon grounded in the physical instantiation of knowledge.}

\begin{figure}[h]
	\centering
	\includegraphics[scale=0.1]{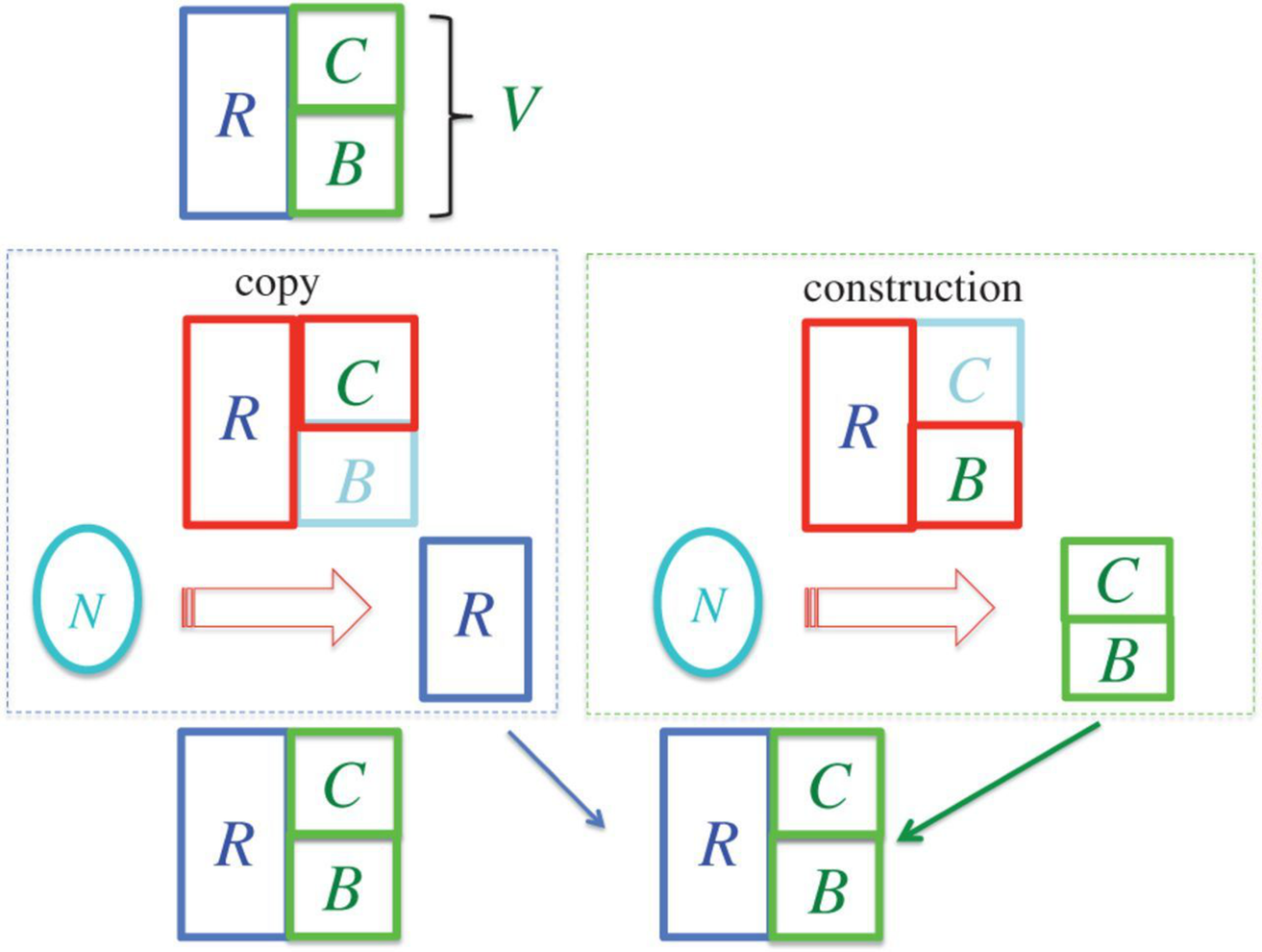} 
	\caption{{The logic of self-reproduction with the replication-vehicle logic.  An accurate self-reproducer (top) consists of the replicator R (in blue) and the vehicle V (in green) containing the copier C and the constructor B. In the copy phase, C copies the replicator R. In the construction phase, B executes the recipe in R to build a vehicle from generic resources N. Bottom: the copy of R and the newly constructed vehicle form the offspring.}}
\end{figure}

{The ultimate development of the constructor theory of life is achieving a theory of the universal constructor. This is a device capable of causing any physical transformation or delivering any physical object that any other device could under the same circumstances, while requiring no input from the user except knowledge – that is to say, a program. Universality for constructors is radically different from that of computers, \cite{DEU25}.}
Unlike their analogues for computation and life, universal constructors are not built out of elementary construction primitives, analogous to bits, qubits and replicators: a universal constructor is necessarily a macroscopic device. With only knowledge from the user, it must be capable of providing its own power and raw materials for the construction, as well as maintenance for itself.
Hence, a universal constructor is an open system, not just in an unphysical approximation like a computer, where we ignore its need for maintenance, power, etc., and inputs.
We expect that a proof exists, from the principles of constructor theory, that a universal constructor can be built by non-universal constructors under no-design laws. The quest for such a proof may also provide insights into why the laws of physics are as they are – perhaps making unnecessary any appeal to the Weak Anthropic Principle to explain the ‘fine-tuning problem’, \cite{DEU25, FINE}. {One possible line of testing concerns the existence of thresholds separating universal constructors from other programmable constructors. If constructor theory is correct, there may exist principled distinctions between systems capable only of narrow task execution and those capable of arbitrarily extensible construction given appropriate information. Programmable nanorobots or cells could be used to test such thresholds and related models. 
More radically, a theory of universal constructors might allow one to identify transformations that are physically possible in principle but inaccessible within standard dynamical descriptions.  If a universal constructor framework predicted new classes of achievable transformations — for example, forms of autonomous error correction, self-repair, or hierarchical construction not anticipated in ordinary physical modelling — then demonstrating such transformations experimentally would provide strong evidence in favour of the constructor-theoretic perspective. Experimental tests arising from such a theory would therefore not merely confirm isolated predictions, but probe whether nature genuinely obeys laws formulated in terms of constructional capability rather than dynamical evolution alone.}

\subsection{The Ultimate Constructor-Theoretic Law of Physics}

We now explore what kind of constructor-based principle(s) could serve as a selector for the ultimate law(s) of physics. Assuming a universal constructor is possible, we are led by the vision that the starting point is the division of all tasks into possible and impossible with respect to the universal constructor. In this sense, a logical model is provided by thermodynamics, in particular Caratheodory's axiomatic formulation of the second law, \cite{CAR, LIE}. This formulation of the second law says that in the vicinity of every thermodynamic state of a system, there are states that are adiabatically inaccessible from it. As we mentioned, in Joule's experiments, the temperature of an adiabatically closed system can be raised by steering it, but it can never be lowered. 

In addition, we are motivated by various observations that the laws of quantum theory are in a number of well-defined ways ``tighter" than the laws of classical physics (Newtonian mechanics, Maxwell's equations, General Relativity). For instance, most non-linear modifications of quantum mechanics (but not all) lead to the violation of causality by allowing us to perfectly discriminate non-orthogonal states \cite{Gisin}. Similarly, for the modification of the Born rule (e.g., by generalising the mod square rule for probabilities) \cite{Aaronson}. It seems that even seemingly innocuous changes of the basic axioms of quantum mechanics lead to violations of relativity or thermodynamics or some other strongly held computational principles (such as the division between the easy and hard tasks in the computational complexity hierarchy, see \cite{vanDam}). 

Finally, we expect the usual dynamical laws of physics, such as the Schr\"odinger equation, to be selected by the ultimate constructor-theoretic principles. In that sense, we would like the constructor principle to do as much heavy lifting as possible, and we would ideally have to rely, if at all, on very few extra assumptions.  

With this in mind, it seems natural to offer the following generalisation of Caratheodory as a potential candidate for the ultimate principle of constructors: in the vicinity of every possible task, there exist tasks that are impossible. 
For this principle to be useful to us, we first need to mathematically define the neighbourhood of a given task. This can be done by formally introducing a distance measure on the set of all tasks (which can be given by the subsidiary theory, see the appendix). {Note that a possible task is not equivalent to an allowed process, so one needs to carefully analyse how to define the neighbourhood of a possible task.} Moreover, we need a clear criterion for what makes a task impossible (namely, which particular principles are used to rule out tasks), \cite{DEUMA25}. As we said, this is frequently done by using either the laws of relativity (such as tasks that allow superluminal communication) or the laws of thermodynamics (tasks that violate the first, second, or third law \cite{Maruyama}). However, it remains an open question for future research whether there is a formal way to constrain tasks in general. This direction is currently a work in progress, and we believe that, in the near future, we will be able to report further new developments. We also expect the constructor-theoretic generalisation of Caratherodory's principle to result in new tests and experimental proposals, for instance, to compare its predictions with current statements of the second law, and the constructor-theoretic expression of the second law itself, \cite{MAT1, MAT2}.

\section{Conclusions}

Testing the laws of Constructor Theory is possible. Recently, a number of exciting experimental proposals have emerged, and experiments have been performed using high-quality photonic and NMR
qubits. In this work, we have reviewed these developments. There are a number of possible exciting avenues ahead:

\begin{itemize}
\item {\bf Post Quantum Information and Computation}\\
Constructor theory's principles could be applied to modified versions of quantum theory (see e.g. non-commuting qubits, \cite{SAM}, hybrid quantum-classical theories, superselected quantum field theories, quantum gravity theories) in order to identify observables and lead to new tests of these alternative laws vs standard quantum theory. 
\item {\bf Quantum Biological Systems (Self-Reproduction and Evolution)}\\
Constructor theory provides the fundamental principles underpinning self-replicating systems (like DNA).
Experiments in synthetic biology could test whether Constructor Theory correctly explains the limits of self-replication and mutation. Possible experiments could include studying non-DNA-based life forms to see if self-replication requires the same fundamental information principles as the familiar form of life.
\item {\bf Thermodynamics}\\
Traditional thermodynamics is statistical, but Constructor Theory aims for exact laws about possible and impossible transformations.
It would be interesting to design new types of thermal machines (such as molecular-scale heat engines or actual biological systems) and check if they follow the predictions of constructor theory in the domains where current thermodynamics laws do not apply because of scale issues. One possibility could be to test the constructor-based irreversibility and potential deviations from the current second laws in qubit-based thermalisation machines. 
\item {\bf Cosmology and quantum gravity}\\
Constructor Theory suggests that physical laws should be rewritten in terms of possible/impossible tasks rather than equations of motion. These constraints can act as selectors of the initial conditions. It would be interesting to study cosmological implications of these principles and scan for potential tests of existing cosmological theories. 
\end{itemize}

Whether or not constructor theory fulfils the grand vision of providing a novel underpinning to supplement dynamical laws, it has already led to new tests and predictions in domains where quantum theory cannot be directly applied. Moreover, its principles can be used as guidelines to guess the future theories of physics and to design experiments which complement existing ones. The current state of theoretical physics lacks experimental inputs that fruitfully challenge the predictions of current theories. {Constructor theory tests
represent an essential addition to fundamental physics research. Similar to GIE, they allow us
to transcend the limitations of current experimental methods and facilitate the consideration
of alternative candidate theories. Ultimately, these tests and alternative theories can inform
the search for a successor to the quantum theory that might unify it with general relativity.}  

We hope and expect that constructors will provide the necessary bridge between the inanimate matter and living systems. The notion of a universal computer is clearly insufficient for this purpose, and the question remains whether constructors have sufficient generality to play this role of ``upgrading" chemical reactions into biological functionality. It would certainly be interesting if such a generalisation of computers provided us with both the impetus for new laws of physics and the unification of chemistry and biology. 

\appendix \section{Appendix}

\subsection{Disambiguation}

Here, we provide a little more detail on comparing the constructor theory with other approaches. 
Generalised Probabilistic Theories (GPTs) \cite{BAR} and Operational Probabilistic Theories (OPTs) \cite{REF33, REF31} aim to formulate information-theoretic principles independently of the standard Hilbert-space formalism of quantum theory. Unlike such frameworks, constructor theory aims at a fully explanatory physical framework not based on primitive probabilistic or measurement-theoretic notions.  In particular, unlike constructor theory, GPTs are intrinsically probabilistic, and OPTs typically take operational primitives such as preparations, transformations, and measurements as fundamental. In that sense, they are not explanatory physical theories in the same way constructor theory is.
 Resource-theoretic frameworks, on the other hand, express existing dynamical laws (not just quantum theory, but also other dynamics); while CT provides novel physical principles (e.g. the interoperability law) which supplement dynamical laws. 
There are also ‘hybrid quantum-classical’ or ‘semiclassical dynamics’ \cite{SUD}, which aim to go beyond quantum theory's formalism. These are different from constructor theory because the latter avoids assuming a specific hybrid or semiclassical dynamics, only general dynamics-independent principles formulated in terms of possible/impossible tasks. Finally, Constructor theory of information and axiomatic reconstructions of quantum theory differ mainly in their aims, starting points, and the role they assign to information. Constructor Theory of Information, is intended as a new foundational framework for physics. Rather than beginning with mathematical structures or probabilistic rules, it describes the world in terms of which physical transformations are possible and which are impossible. Within this framework, information is defined through the possibility of tasks such as copying or distinguishing states, making it an explicitly physical notion grounded in counterfactual possibilities.
By contrast, Axiomatic reconstructions of quantum theory— \cite{REF32} —aim to recover the formalism of quantum theory from a small set of clear, physically or informationally motivated principles. These approaches typically take notions such as probability, measurement, and information-processing constraints as primitives, and show how the structure of quantum mechanics follows from them. The key distinction is therefore that constructor theory seeks to go beyond quantum theory by providing a deeper, more general account of physical law in which information is fundamental, whereas axiomatic reconstructions take quantum theory itself as the object to be explained. In constructor theory, quantum theory is only one of the dynamics compatible with the super-information axioms, while in axiomatic approaches it is singled out as the endpoint of the axioms.

\subsection{A quantum-theory model for Possible Tasks}

We now present a quantum model for a constructor, following \cite{MAT1, MAT2}. Consider a composite system consisting of two quantum subsystems, $C$ and $S$, with a total Hilbert space given by ${\cal H}={\cal H}_C\otimes {\cal H}_S$. Let $U$ be a fixed unitary operator describing their interaction. We define $\Sigma(X)$ as the +1-eigenspace of the projector $X$. 

Given also any general operator $B$, define $B^{(C)}=B\otimes \identity$ and $B^{(S)}= \identity\otimes B$.

Consider two attributes of $S$, defined as ${\bf x}=\Sigma(X^{(S)})$ and ${\bf y}=\Sigma(Y^{(S)})$, where $X$ and $Y$ are two orthogonal projectors. Each attribute represents the set of states of $S$ in which the corresponding projector is sharp with value 1. Define a task $T= \{{\bf x} \rightarrow {\bf y}\}$.

{Next, define the set:}
\begin{equation}
V_T=\{\ket{\psi}\in {\cal H}_C\;|\;\forall \ket{x}\in \Sigma(X^{(S)})\;,\;U(\ket{x}\otimes\ket{\psi})\in \Sigma({Y^{(S)}})\}.
\end{equation}

{It is straightforward to verify that $V_T$ is a vector space. The normalised, non-zero elements in $V_T$ correspond to states of $C$ such that, when $C$ is initialised in one of these states and interacts with $S$ in state $\ket{x} \in \Sigma(X^{(S)})$ with attribute ${\bf x}$, it transforms $S$ into a state possessing attribute ${\bf y}$. Note that in the final state, $C$ and $S$ can become entangled, and $C$ may lose its ability to repeatedly perform the transformation. A necessary set of conditions for $C$ to function as a constructor for task $T$ are the following:}
\begin{itemize}
\item {The subspace $V_T$ includes non-zero elements};
\item There exists a non-trivial subspace $W_T\subseteq V_T$ such that
\begin{equation}
U\left (W_T\otimes \Sigma(X^{(S)})\right )\subseteq W_T\otimes \Sigma(Y^{(S)}).
\end{equation}
This condition ensures that the states in $W_T$ retain their ability to implement the transformation $T$ indefinitely. Denote by $\Pi_{W_T}$ the orthogonal projector in the smallest subspace $W_T \subseteq {\cal H_C}$ that meets this condition.
\end{itemize}

If these conditions hold, we can define the (non-empty) set  such that:
\begin{equation}
V_{C_T}=\{{\psi}\in {\cal H}_C \;|\;\forall {x}\in \Sigma({X^{(S)}})\;,\;U(x\otimes{\psi})\subseteq {{{W_T}^{(C)}}\otimes \Sigma(Y^{(S)}}\},
\end{equation}
which is also a vector space. {Norm-1 vectors in this subspace are physical states that either belong to $W_T$ or are mapped into it after a single application of $U$.} The projector $\Pi_{C_T}$ onto this subspace serves as the {projector for being a constructor} for the task $T$.
The possibility of the task $T$ in quantum theory implies the existence of a subspace $V_{C_T}$ satisfying these conditions.

\subsection{Distinguishability as a physical property}

We express more formally the conditions for distinguishability expressed in the constructor theory of information, following \cite{DEUMA, MAP}.

Define the cloning task for a variable $X$ as:
\begin{equation}
\label{eq1}
C(X)\doteq\bigcup_{{x}\in X}\left\{({\bf x},{\bf x_0})\rightarrow ({\bf x},{\bf x}) \right\};
\end{equation}

\noindent where $\bigcup$ denotes set union, and $x_0$ is a fixed attribute. The set $X$ is said to be copiable if the task $C(X)$ is possible for some $x_0$. In quantum theory, this task is feasible if and only if all elements of $X$ are mutually orthogonal; otherwise, it is impossible. For instance, if $X$ is a Boolean variable, $X={{\bf 0},{\bf 1}}$, and ${\bf x_0}={\bf 0}$, then the task $C(X)$ can be implemented via a controlled-NOT gate. This is a generalisation of the cloning task in quantum theory.

An information medium is a substrate where both the cloning task $C(X)$ and the permutation task

\begin{equation}
\label{eq2}
\Pi(X)\doteq \bigcup_{{x}\in X}\left\{{\bf x}\rightarrow \Pi({\bf x}) \right\};,
\end{equation}

\noindent are possible for all permutations $\Pi$ on the labels of the attributes in $X$, for some attribute ${\bf x_0}\in X$. In quantum theory, a set of orthogonal states without additional symmetries or superselection rules qualifies as an information variable.

\noindent The task $C(X)$ corresponds to {\sl copying}, or cloning, the attributes of the first substrate onto the second, target substrate, while $\Pi(X)$ represents a logically reversible computation for a given $\Pi$. For example, a qubit is an information medium when equipped with any two orthogonal quantum states, $X={{\bf 0}, {\bf 1}}$.

A variable $Y$ is said to be {\sl distinguishable} if the task
\begin{equation}
\label{eq3}
\bigcup_{{y}\in Y}\left\{{\bf y}\rightarrow {\bf q_y} \right\}
\end{equation}

\noindent is possible, where the variable ${{\bf q_y}}$, having the same cardinality as $Y$, is an information variable.
Remarkably, this definition is non-circular (it does not presuppose any pre-defined notion of information), scale- and dynamics- independent, and non-probabilistic. 

{\bf Acknowledgements} \;\; The authors are grateful to Artur Ekert, Giuseppe Di Pietra, Nicetu Tibau Vidal, Simone Rijavec and Maria Violaris for comments and criticism on earlier versions of this manuscript. This publication was made possible through the support of the Gordon and Betty Moore Foundation, the Eutopia Foundation, B. Vass, and the Conjecture Institute.

\end{document}